\documentstyle[epsf]{ICSM}
\font\tenmsam=msam10
\def\gaeq{{\hbox{\tenmsam\symbol{"26}}}}
\def\laeq{{\hbox{\tenmsam\symbol{"2E}}}}
\begin{document}
\title{Organic Superconductors: Reduced Dimensionality and Correlation
Effects  } 
\author{C. Bourbonnais\\
\it Centre de Recherche en Physique du Solide, Universit\'e de Sherbrooke,
Sherbrooke, Qu\'ebec, J1K-2R1}

\abstract{In this tutorial we will tackle  the problem of  electronic correlations
in quasi-one-dimensional organic superconductors. We will go through different 
pieces of experimental evidence showing the range of applicability of the  Fermi and
Luttinger liquid descriptions of  the normal phase   of the Bechgaard salts series
and their sulfur analogs. 
\\~\\ {\it Keywords }: organic superconductors, many-body  theories,  magnetic measurements,
magnetic phase transitions. }

\maketitle

\section{Introduction}
In solid-state physics the traditional approach to  understand ordinary metals
relies on the  Landau theory of Fermi liquid. The concept of quasi-particles 
put forth  by Landau proved to be quite successful in describing    low-energy
excitations,  which effectively  behave like non-interacting electrons.  Besides
the possibility of long range order which limits its range of applicability,
the Fermi liquid theory turns out to be highly sensitive to the
spatial dimensionality of the electron system.  This is especially  true  for purely
one-dimensional systems  where the Landau  Fermi liquid theory breaks down. It is
pretty well understood that in one dimension there are no single-particle states 
that adiabatically connect with the non-interacting electrons.  Instead elementary
excitations are made of collective modes of bosonic character \cite{1D}.

 The possiblity that such   non-Fermi liquid  correlation effects   may be relevant
to the description  of the normal phase of real materials like quasi-one-dimensional
organic conductors has attracted   lot of interest over  the past two decades
\cite{Jerome}.  However, since  these systems are not strictly one-dimensional, this
raises the delicate  question of the extent to which Fermi and non-Fermi liquid
concepts are relevant to the understanding of the normal phase of  these molecular
compounds.
In this tutorial we will pursue this issue for the well known isostructural  series
of sulfur-based (TMTTF)$_2$X and selenium-based (Bechgaard salts) (TMTSF)$_2$X 
organic conductors. 
    We will review  recent progress made in the analysis of  the   NMR spin-lattice
relaxation rate, magnetic susceptibility and transport  measurements in connection
with correlation effects and non-Fermi liquid features 
  found in these  strongly anisotropic materials.

\section{The low-field phase diagram}
 Following the synthesis of the isostructural  sulfur
(TMTTF)$_2$X and selenide (TMTSF)$_2$X series of 
quasi-1D  organic conductors, it was soon recognized that   a remarkable
continuity can be drawn  
whenever their phase diagrams are combined as a function of
applied hydrostatic pressure  or anion (X= PF$_6$,
AsF$_6$, Br... ) substitution.  Depending on the choice of the anion  at ambient
pressure,
members of the sulfur series can develop   either   spin-Peierls or 
 commensurate/localized  antiferromagnetic long range order whereas  either
itinerant antiferromagnetism (SDW) or superconductivity can be found for
the selenide series under  similar pressure conditions. Under  pressure, however,
the spin-Peierls and localized antiferromagnetic orderings alternately become 
unstable to the benefit  of phases found in  selenides thereby  ensuring
the   characteristic sequence SP$ \to$ AF $\to $ S of transitions shown in Figure~1.

 \begin{figure}[htb] 
\epsfxsize0.95\hsize
\epsffile{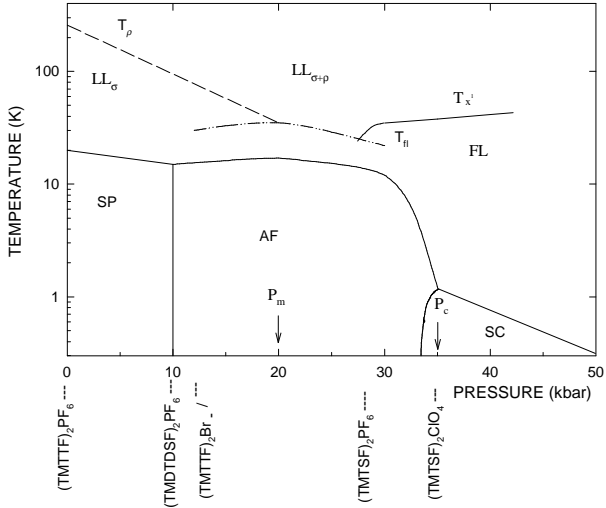}
\caption{ Phase diagram of the of (TMTTF)$_2$X and (TMTSF)$_2$X under pressure.}    
\label{diagramme}   
\end{figure}

The survey of existing results at low magnetic fields reveals  that the critical
temperature bracket for AF ordering  is $T_N
\approx1\ldots 24$K ( the highest $T_N$ being found in sulfur series)  
and  $T_{SP}\approx  8\ldots 19$K for the spin-Peierls ordering,   while for
supeconductivity, it is scarecely more than 1K.

The study  of the normal phase has shed some special light on the
central role played by antiferromagnetism in the phase diagram of these 
quasi-1D materials. This is particularly  revealing for (TMTTF)$_2$X
compounds for which  precursors to  magnetic ordering become 
manifest  at a characteristic temperature 
$T_\rho$ that  can be in magnitude  far  
  above  $T_N$  under  low pressure conditions. The  scale
$T_\rho$  signals the loss of metallic character 
and in most cases the thermal activation of  carriers  corresponding to a gap
$\Delta_\rho\approx 2\ldots 3 T_\rho$ in  the normal phase
temperature dependent electrical resistivity (Figure~2).

 \begin{figure}[htb] 
\epsfxsize0.95\hsize
\epsffile{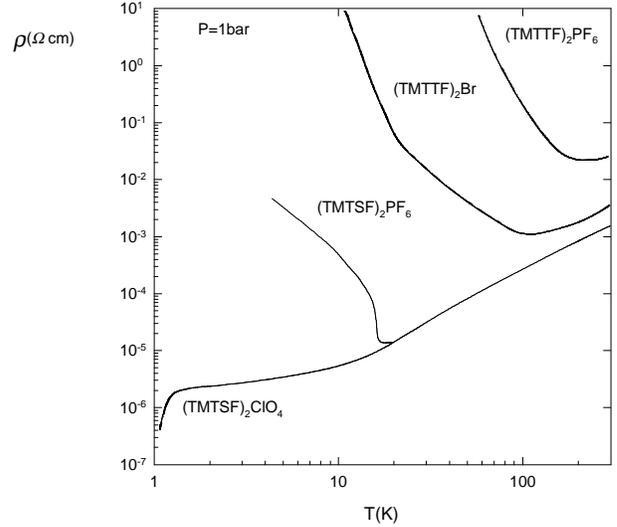}
\caption{ The temperature dependence of the resistivity for representatives of the Bechgaard salts and
their sulfur analogs.}    
\label{Resistance}   
\end{figure}

 This contrasts with  the temperature 
variation of magnetic
susceptibility shown in Figure~3,  which indicates that long wavelength  spin
degrees of freedom remain unaffected at
$T_\rho$. As first pointed out by 
Emery {	\it et al.} \cite{Emery82}, this is
the signature of a  ``4$k_F$" Mott-Hubbard localization of the carriers whose origin
would result from  1D many-body physics at half-filling  whenever
repulsive interactions among carriers are combined with   a  slight
dimerization  of the organic stacks and  the  $4k_F $ anion potential. Here the
possibility of sizeable short range repulsive  Coulomb interaction in these kinds of
molecular complexes  is confrimed by detailed quantum chemistry calculations
which tell us that the bare one-site repulsion parameter $U$  is of the order of one
ev
\cite{Ducasse}.   

\begin{figure}[htb] 
\epsfxsize0.95\hsize
\epsffile{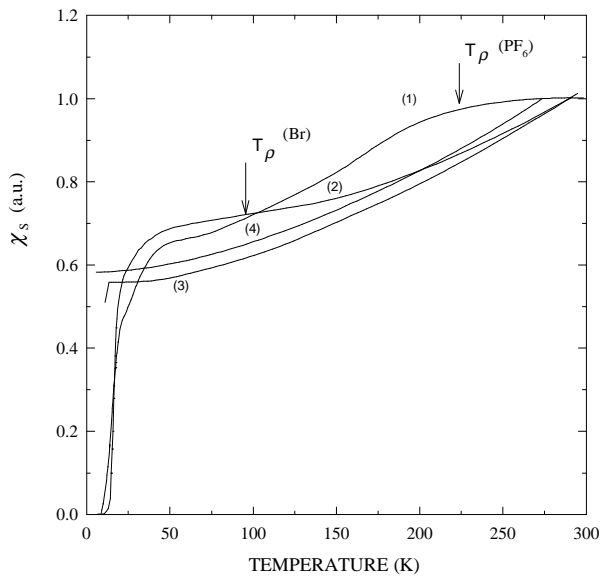}
\caption{Temperature dependence of the magnetic susceptibility of (TMTTF)$_2$PF$_6$
(1), (TMTTF)$_2$Br (2), (TMTSF)$_2$PF$_6$ (3) and (TMTSF)$_2$ClO$_4$ (4) at ambient
pressure. }    
\label{Chis(T)}   
\end{figure}

\subsection{ The Luttinger liquid concept and the description of the normal phase }
The Mott-Hubbard instability is   
 symptomatic of the  breakdown of the  Fermi liquid picture  of
the normal phase whose description  would rather correspond  to a limiting form
of a {\it
Luttinger liquid}. The concept of a  Luttinger liquid    was   coined by
Haldane fifteen years ago in order  to characterize the  peculiar
features  displayed by a      many-electron quantum system in 1D
\cite{Haldane}. It has been  known for almost three decades that unlike a Fermi
liquid, a Luttinger liquid at  zero temperature has  neither quasi-particle
states  nor a Fermi distribution step at the Fermi level \cite{Mattis}. The
instability of single particle states occurs  at the expense of  collective
 modes related  
 to spin and charge acoustic branches of excitations.  
These modes of different velocities $v_\sigma$ and $v_\rho$  are  decoupled
at large distance  and  lead to the so-called {\it spin-charge separation}.
At half-filling, the Mott-Hubbard localization magnifies the spin-charge
separation in such a way that the gap $\Delta_\rho$ imposes a finite 
coherence length $\xi_\rho \sim  v_\rho/T_\rho$ for spatial
correlations of charge degrees of freedom, while the coherence length
$\xi_\sigma\sim v_\sigma/T$  for spins is unaltered  and reaches   infinite
range in the low temperature limit. 

These features, however consistent with
what is commonly observed in sulfur series at low pressure, are not the only
 signs in favor of a Luttinger liquid picture in these systems. 
Elaborate calculations using the abelian bosonization technique, predict that
the single particle spectral weight
$A(k,\omega)$ (which gives the probability of having a single particle state
at the wavevector $k$ with an energy $\omega
$ measured from the Fermi level $(\hbar=1)$) has no single particle 
peaks  but rather shows branch-cut singularities  with  non-universal
exponents \cite{Voit}. Albeit photoemission experiments have been carried out
recently on a member of the selenide series \cite{Dardel}, direct measurements of
one-particle spectral properties of sulfur compounds are missing so far.
 Hallmarks of the Luttinger liquid  can also be found  in  two-particle response
functions, which turn out to be  more accessible experimentally. This is especially
true for the temperature dependent antiferromagnetic ($2k_F$) susceptibilty
$\chi(2k_F,T)$, which can be extracted from NMR spin-Lattice relaxation experiments. For 
repulsive interactions, $\chi(2k_F,T) \sim T^{-\gamma}$ develops   
 a power law singularity with 
 an exponent $\gamma= 1-K_\rho$ that  is
expressed in terms of the non-universal charge degrees of freedom index $K_\rho$
whenever  magnetic anisotropy is irrelevant. The Mott-Hubbard
localization of   charge degrees 
of freedom at $T<T_\rho$ imposes the exact constraint $K_\rho=0$, which leads to  
$\chi(2k_F,T) \sim T^{-1}$ and  which actually conveys the same type of singularity 
found in   the exactly solved isotropic Heisenberg model \cite{Voit}.

\begin{figure}[htb] 
\epsfxsize0.95\hsize
\epsffile{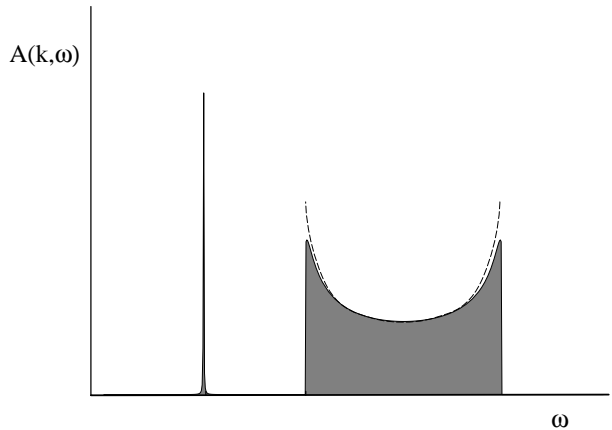}
\caption{Schematic one-particle spectral weight  of a Luttinger liquid for the one-dimensional case
(dashed line). In the quasi-one-dimensional case, the single-particle dimensionality
crossover introduces a quasi-particle peak and the  smearing of   the power-law
singularities. }    
\label{Poids}   
\end{figure}

\subsection{Nuclear relaxation as a probe for Luttinger liquid}

In the past fifteen years or so, a great deal of interest has been  devoted to the
study of electronic correlations using   NMR spin-lattice relaxation rate
($T_1^{-1}$) measurements.  The familiar Moriya formula
\begin{equation}
T_1^{-1} \propto T \int d^Dq {{\rm Im}\chi({\bf q},\omega)\over \omega}, 
\end{equation}
that connects relaxation to the imaginary part of the dynamic spin susceptibility   
actually tells us how a local probe like  NMR   makes 
$T_1^{-1}$
 sensitive to static, dynamic {\it and} dimensionality $D$ of spin fluctuations  for
all wavevectors
${\bf q}$ \cite{Bourbon89}. From the results of Ref.\cite{Bourbon89,Bourbon93}, the
exact $D=1$ Luttinger liquid prediction for $T_1^{-1}$ takes the form:
\begin{equation}
T_1^{-1}= \ C_0T\chi_s^2(T) + C_1 T^{1-\gamma},
\label{T1}
\end{equation}   
which consists of   contributions coming  from uniform ($q= 0$) and
antiferromagnetic ($q\approx  2k_F$) spin fluctuations. Both 
generate deviations with respect to the Korringa law $(T_1T)^{-1}=$ constant, 
 found in ordinary  metals with weak interactions.  The uniform contribution, albeit 
related to the non-singular temperature dependent magnetic  spin susceptibility,
dominates at  high temperature while the antiferromagnetic  power law enhancement
$(T_1T)^{-1}\sim C_1T^{-\gamma}$ in principle  allows 
 to  extract the  Luttinger liquid exponent $\gamma$. The  combination of 
$\chi_s(T)$ and
$T_1^{-1}$ measurements  then proves to be  a powerful tool in  the analysis of
correlation effects in  the normal state of  low-dimensional organic conductors
\cite{Bourbon89,Bourbon93,Wzietek93}.

\begin{figure}[htb] 
\epsfxsize0.95\hsize
\epsffile{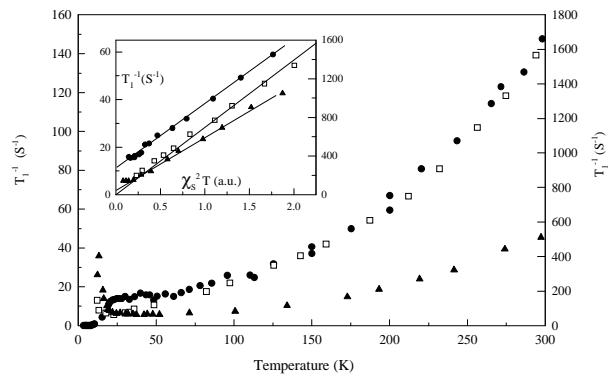}
\caption{$T_1^{-1}$ {\it vs} $ T$ for (TMTTF)$_2$PF$_6$ (circles, $^{13}$C, left scale),  (TMTTF)$_2$Br
(triangles, $^{13}$C, left scale) and (TMTSF)$_2$PF$_6$ (open squares, $^{77}$Se, right scale) at $P$= 1bar. In
the inset
$T_1^{-1}$ {\it vs} $T\chi_s^2$ in the normal phase using the data of Figure 3. After
Refs. [10,11].}    
\label{test}   
\end{figure}

 This ability has been 
remarkably illustrated in the case of (TMTTF)$_2$X compounds  as  typified
by  the  selected results shown in  Figure~5.  Let us
consider for example  the normal state  of the spin-Peierls system (TMTTF)$_2$PF$_6$ 
at ambient pressure. This compound presents a well defined  resistivity  minimum at
$T_\rho\approx 220$K \cite{Coulon}, whereas the spin susceptibility $\chi_s(T)$ 
shown in Figure~3,
produces   a regular downward slope as the temperature is lowered. As for the 
$T_1^{-1}$ temperature profile  
 given in  Figure~5, it also presents a monotonic decrease with   an
upward curvature down to the low  temperature region  where $T_1^{-1}$ shows signs of
saturation.  This persists down to 40K or so, below which the system is
dominated by strong spin-Peierls lattice correlations, as  borne out by x-ray 
experiments  \cite{Pouget}. Plotting   
$T_1^{-1}$ {\sl vs}
$T\chi_s^2(T)$  in the temperature  domain above 40K ( where  lattice
fluctuations are weak),  one finds that the Luttinger liquid prediction (\ref{T1}) 
with 
$\gamma=1 (K_\rho=0)$ is rigourously obeyed (inset of Figure~5). Therefore the
existence of a constant term in $T_1^{-1}$  over all the temperature domain
considered soundly supports the existence of a Mott-Hubbard localization which
essentially coincides to a Luttinger liquid with a charge   gap and gapless spin
excitations. Similar features  have been tracked down in other members of the same 
series. In this connection we have reproduced  in  Figure~5, the
 $T_1^{-1}$ temperature profile obtained for  
(TMTTF)$_2$Br where
$T_\rho
\sim 100$K   and $T_N \approx 15$K \cite{Wzietek93}. Using the susceptibility
data of Figure~3, the  corresponding 
$T_1^{-1}$ {\sl vs}
$T\chi_s^2(T)$ plot  in the inset shows that in the normal phase ( above  the
critical domain taking  place below 30K), the expression (\ref{T1}) with
$\gamma=1$ is again fully obeyed but with $C_1$ being smaller \cite{Wzietek93}.

\subsection{Interchain coupling   and the one- and two-particle 
instability  of the Luttinger liquid }

 Given the quasi-1D  character  of materials like (TMTTF)$_2$X, the presence of long range ordering invariably
indicates that the Luttinger liquid becomes unstable at low temperature. It can be
shown,  however, that it can 
 still influence the formation of critical ordering  in this case antiferromagnetic.
The microscopic source of instability of the Luttinger liquid comes from the
interchain coupling and in particular the interchain single particle hopping
$t_\perp$ \cite{Firsov,BC1,B2}.  This coupling                               is  essential to the
 propagation  of antiferromagnatic correlations  perpendicular to the chains. According to band
calculations \cite{Grant}, however, 
$t_\perp \ \laeq\  100$K in  sulfur compounds, which means that a coherent
interchain   single particle hopping (band motion) is  made irrelevant owing to the
presence of  a large charge  gap
$\Delta_\rho > t_\perp$ in the normal phase \cite{Brazo}. Nevertheless, bound electron-hole pairs can tunnel to
neighboring chains via
 the virtual hops of single carriers leading to a perpendicular antiferromagnetic  
kinetic exchange  coupling  $J_\perp \approx 2\pi v_F t_\perp^{*2}/ \Delta_\rho^2$
(where
$t_\perp^*$ stands as a renormalized value of $t_\perp$). The critical
 ordering temperature $T_N$ is obtained by considering the  effect of $J_\perp $ in molecular field
approximation  whereas  intrachain correlations are treated exactly. One then gets the Stoner criteria
$1-J_\perp
\chi(2k_F,T_N)=0$ for $T_N$. Now using the exact power law expression $\chi(2k_F,T) \sim T^{-1}$ at $T\ll
T_\rho$, one readily finds  
\begin{equation}
T_N \approx t_\perp^{*2}/\Delta_\rho, 
\label{strong}
\end{equation} 
which {\it increases} as $\Delta_\rho$ or $T_\rho$ {\it decreases} under pressure.
$T_\rho$ will merge in the critical domain of the transition and  becomes
irrelevant.  In this range of intermediate pressure, the normal phase is metallic
corresponding to a Luttinger liquid with  gapless  spin and charge excitations (Figure~1). As
for the  interchain exchange interaction, it takes the form $J_\perp
\approx 2\pi v_F \tilde{g}^{*2} t_\perp^{*2}/T^2$, where $\tilde{g}$ is a normalized  effective
electron-electron coupling constant divided by $\pi v_F$. Therefore whenever 
$T> T_{x^1} \approx t_\perp^*/2$ ( that is above the single-particle dimensionality crossover temperature that
marks the thermal deconfinement of  along the stacks), $J_\perp$ turns
out  to be still active in the development of long range order. The Stoner criteria
in this case leads to 
\begin{equation}
T_N \approx 2\tilde{g}^*T_{x^1} ,
\label{weak}
\end{equation}   
which {\it decreases} through $\tilde{g}^*$ under pressure \cite{BC1,B2}. Ultimately, the
$T_N$ pressure profile  will show a maximum at a critical pressure $P_m$
where
$T_\rho$ merges into the critical domain.  A maximum of this kind has been found in
all sulfur compounds that were the  subject of pressure studies.   The results of
Klemme {\it et al.}\cite{Klemme} for the (TMTTF)$_2$Br, which are reproduced  in 
  Figure~6, indeed illustrate  how the   
interchain exchange  coupled to the  Luttinger liquids of isolated chains give a controlled description of 
 the antiferromagnetic transition in sulfur compounds up to intermediate
pressure.  

\begin{figure}[htb] 
\epsfxsize0.95\hsize
\epsffile{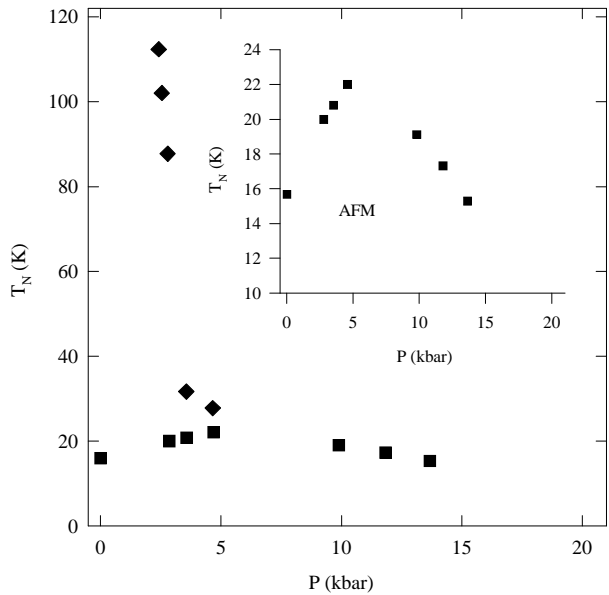}
\caption{ $T_N$ as a function of  pressure for (TMTTF)$_2$Br. In the inset the
pressure profile of $T_N$ near $P_m$. After Klemme {\it et al.,} Ref.[19].}    
\label{TNBr}   
\end{figure}

The interchain kinetic exchange will keep its key role for long range ordering  as
long as the carriers stay thermally confined in the transverse direction. In fact, as
one moves away from   the maximum of  $T_N$  by increasing  pressure,
the downward renormalization  of $t_\perp^*$ gets
weaker due to a reduction of intrachain  many-body effects. This progressively
raises the single-particle dimensionality crossover  temperature
$T_{x^1}$   up to a point where the
exchange-induced temperature scale for critical fluctuations
$T_{fl} \approx 2T_N$, as obtained from (4),  becomes smaller than
$ T_{x^1}$ \cite{Note}. This range of pressure
 then shows  electronic deconfinement which is  a one-particle type of instability
of  the Luttinger liquid that announces the  
formation  of a {\it Fermi liquid} component  \cite{BC1,B2,note2}. 
It follows that the two-dimensional character of nesting properties  at  the
wavector ${\bf Q}_0=(2k_F,a\pi)$ picks up  some coherence
 below $T_{x^1}$ which can  no longer be neglected in the
evaluation of the critical temperature $T_N$. Here the modulation wavevector ${\bf
Q}_0$ is found to be incommensurate for  the sulfur compounds at high
pressure
 and in the selenides as well where ($a\approx .24$)
\cite{Klemme,Takahashi}. From the point of view of  perturbation theory, such a
coherence    marks a progressive decoupling of  infrared singularities  pertaining to
the ``2$k_F$" density-wave and superconducting Cooper channels whose interference is
at the heart of   the existence of  a Luttinger liquid  in one dimension. Thus for
those  degrees of freedom involved in the Fermi liquid component    below $T_{x^1}$, 
the elementary logarithmic infrared singularity of the spin-density-wave channel,
which reads 
$\chi^0({\bf Q}_0,T)\approx (2\pi v_F)^{-1} \ln(T_{x^1}/T)$ for perfect nesting
conditions,   may be singled out and  summed up at the one-loop
 ladder level of the perturbation theory. This   
yields  the  Stoner criteria
$1-\lambda^*\chi^0({\bf Q}_0,T_N)=0$, from which one obtains
\begin{equation}
T_N \approx T_{x^1}\  e^{-2/\tilde{\lambda}^*},
\end{equation}  
where $\tilde{\lambda}^*$ stands as an  effective coupling constant that
combines  the renormalized intrachain interactions {\it and} the  interchain 
exchange at the energy scale $ T_{x^1}$ \cite{BC1,B2}.  Nesting properties below
$T_{x^1}$, however, 
 rely on the electron-hole symmetry relation 
$\epsilon({\bf k})=  -\epsilon({\bf k} + {\bf Q}_0)$ 
at ${\bf Q}_0$, which turns out to be vulnerable to small perturbations that are 
present in the actual spectrum $\epsilon({\bf k})$.  Next-nearest-neighbor
interchain hopping
$t_{\perp,2}$ is  often used to parametrize such an effect that cuts off the
logarithmic singularity in $\chi^0({\bf Q}_0,T)$,  thereby reducing
$T_N$. A perturbation like $t_{\perp,2}$ is  quite sensitive to pressure and  only few kilobars 
are needed to bring
$T_N$ down to zero. 

It is now generally admitted that this simple mechanism is responsible for the 
frustration of the incommensurate spin-density-wave phase found  in selenides
(sulfur) series at low (high)  pressure \cite{Yamaji,Seidel}.  It should be stressed,
however, that the presence of nesting frustration  indicates that a Fermi liquid
component  is present in the normal phase thus confirming that the characteristic
scale $T_{x^1}$ for deconfinement   really  emerges in a  pressure range that  can
safely be located somewhere between  $P_m$  and $P_c$  (Figure~1). 
Since transients are  likely to be associated with such  a crossover, the
deconfinement is not sharply defined so that at a given pressure it should be considered as somewhat spreaded
in temperature. For systems like (TMTSF)$_2$X
at ambient pressure or (TMTTF)$_2$X at relatively high pressure, the 
  accepted range of values  for 
$T_{x^1}$  may differ a lot from author to author. Following the example of what has 
been done for sulfur compounds at low pressure it is thus  preferable to look at 
experiments in order to detect deviations to the  Fermi
liquid  picture  in the normal phase.  If one  looks  for example at
the  existing results  for the temperature dependent $T_1^{-1}$ and
$\chi_s$   in   the ambient pressure metallic phase of (TMTSF)$_2$PF$_6$ ($T_N
\approx 12$K), one observes from  Figure~5 that  deviations
with respect to  the 
$T\chi_s^2$  law  become perceptible below
$T\chi_s^2\approx 1$, which corresponds to the actual temperature  scale $T < 200$K.
According to (\ref{T1}), one can then extract   the
antiferromagnetic part of the enhancement in   $(T_1T)^{-1}$ which is found to have increased by a 
factor 5  at $T\approx 50$K, namely well outside the critical domain of the transition.
This non-critical enhancement  is thus too large to result from  Fermi liquid
conditions in which case  
$(T_1T)^{-1}[{\bf Q}_0]\sim 
$ constant   and should follow a Korringa law. 
This brings us to infer that  a Luttinger liquid picture of the normal phase 
 would still persist at  50K in  (TMTSF)$_2$PF$_6$ at
ambient pressure  suggesting  an upper bound value 
$T_{x^1} < 50$K for  deconfinement. It is worth noting here that a similar enhancement
has been detected in (TMTTF)$_2$Br at 13 kbar
\cite{Wzietek93}.

 A connection has been made  between  these correlation effects and the
anomalous photoemission data of Dardel {\it et al.} \cite{Dardel}, which reveal
a strong depression of the quasi-particle weight at the Fermi level  for
(TMTSF)$_2$PF$_6$ at 50K.  Deviations from the Drude response in optical
conductivity  of the normal phase of these materials have also been related to the
presence of antiferromagnetic correlations, though their influence on 
transport properties  is  in general  poorly understood  \cite{Gruner}.       
  
\section{Confinement under magnetic field }
 
    When a  transverse magnetic field is applied  in the
pressure domain $P \ \gaeq\  P_c$ for compounds like (TMTSF)$_2$PF$_6$ and 
TMTSF$_2$ClO$_4$,  another  sector of  the phase diagram opens up  for which  the
concept of the Luttinger liquid may be relevant to the description of the
normal phase.  When the  magnetic field is increased along the c$^*$ direction, 
this prompts an anomalously large longitudinal resistance and  a upturn in
resistivity at a  temperature $T_\rho(H)$. As shown in Figure~7   for
the (TMTTF)$_2$PF$_6$ under 8.5 kbar and field up to 12T\cite{Balicas}, the
resistivity minimum   moves away from 
 the critical line of  the cascade
of field-induced-spin-density-wave phases. This giant 
magnetoresistance was soon recognized as one of the anomalous feature of the normal phase of these materials
under field\ that does not entirely conform  to  classical magnetotransport
\cite{JeromeSchulz}.

\begin{figure}[htb] 
\epsfxsize0.95\hsize
\epsffile{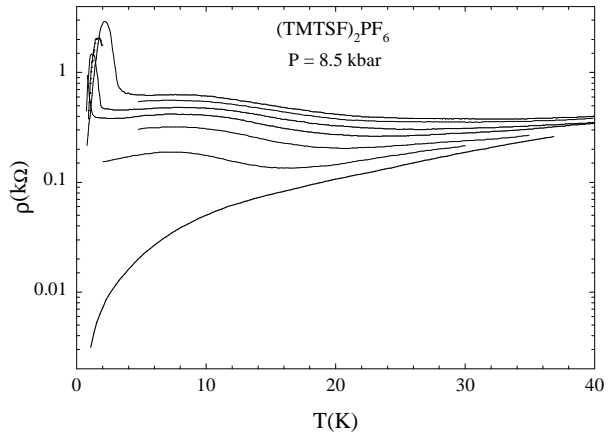}
\caption{ Longitudinal resistance of (TMTSF)$_2$PF$_6$  at $P=8.5$ kbar as a
function of temperature and magnetic field $H= 0, 2, 3, 6, 8, 11, 12.5$T. After Ref.~[27]. }    
\label{rho(T,H)}   
\end{figure}

It was recently proposed that the unidimensionalization of electron motion under
field not only bypasses the effect of
$t_{\perp,2}$ (in inducing perfect but quantized nesting conditions that are
responsible for the cascade of  FISDW  phases \cite{Gorkov}), but improves
longitudinal nesting conditions  at
$(2k_F,0)$ as well. This  gradually  reactivates the singular flow of one-dimensional
many-body effects below
$T_{x^1}$\cite{Behnia}.  Since the dimerization of the organic stacks, though weak,
makes  longitudinal electron-electron umklapp scattering relevant (~the
$g_3$ coupling in the ``g-ology" notation~), 
it may lead to a singular growth of the electron-electron scattering rate which
takes the form
$\tau_e^{-1} \sim T[g_3(T)]^2$.    This may cause an   upturn
 in resistivity which  clearly  is reminescent of the one found in sulfur
series at low pressure (Figure~2) \cite{Gorkov2}.  At variance with the more one-dimensional
sulfur compounds, however, the restoration of longitudinal nesting is only  gradual
under a magnetic field,
 precluding  the
formation of a sharp charge gap that  would entirely destroy  the Fermi liquid
component of the normal phase. The   restoration of the Luttinger liquid  under a  
field, though gradual,  should have  important consequences on spin correlations,
especially those at $2k_F$.  Figure~8 shows $^{77}$Se $T_1^{-1} vs \ T$ and the 
Knight shift $K_s$ ($TK_s^2
\propto T\chi_s^2 $) data of TMTSF$_2$ClO$_4$ at ambient pressure and for two
different fields ( $H= 6$T, with
$T_\rho
\approx  5$K and
$H= 15$T, with
$T_\rho\approx 15$K) \cite{Behnia,Berthier}.

\begin{figure}[htb] 
\epsfxsize0.95\hsize
\epsffile{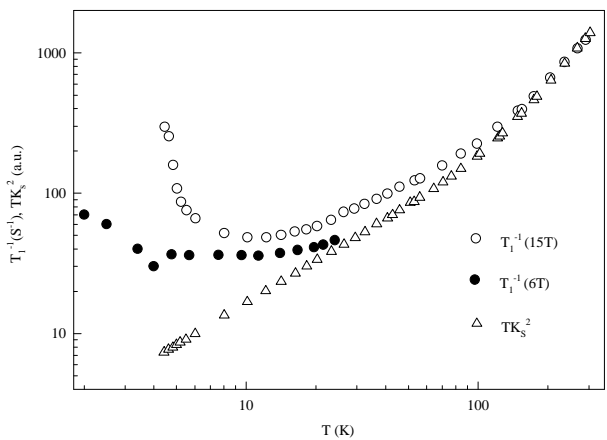}
\caption{Temperature dependence of $^{77}$Se  nuclear relaxation  rate  and Knight shift  for 
(TMTSF)$_2$ClO$_4$ under  magnetic field along c$^*$ at ambient pressure. After Refs. [29,30]. }    
\label{T1H(T)}   
\end{figure}

 As one can see  from the Figure~8, $T_1^{-1}$ is dominated
by the uniform $TK_s^2$ component at sufficiently high temperature but
deviations  connected to the antiferromagnetic part of the relaxation emerge at
low temperature. This enhancement appears to be quite distinct from the
critical $T_1^{-1}$ growth  near the  FISDW transitions taking place at $T_N \approx 2.8
$K (6T) and  4.4 K (15T). Following the analysis made in Ref. \cite{Behnia}, the
relaxation rate in the normal phase can be very well described by an expression of
the form
\begin{equation}
T_1^{-1} \approx  \ C_0T\chi_s^2(T) + C_1(H),
\end{equation}    
where  $C_1(H)$ is a temperature independent contribution that increases as 
$T_\rho(H)$ under a magnetic field.    This is quite similar to  what is
found in the sulfur compounds at higher temperature (see Figure~5). These results
suggest that the progressive unidimensionalization of low-energy states under a field
may gradually reactivate  Luttinger liquid
features in the normal phase precursor to the cascade of FISDW.

Ultimately, this
emphasizes once again 
 that the selenium series and their sulfur analogs are closely
related to one another.    

\medskip\noindent
{\bf Acknowledgments} I would like to thank  my colleagues in Orsay, Grenoble and 
Sherbrooke for their close collaboration on various aspects of this work.

%\thebibliography

%(1993).
%\endthebibliography

\begin{thebibliography}{99}
\bibitem{Note}  $T_{fl}$ actually stands as a two-particle crossover
temperature for the Luttinger liquid which is noted as $T_{x^2}$ in Ref.\cite{BC1,B2}.
\bibitem{note2} Here we adopt a description of dimensionality  crossovers that
neglects transients which can be important for fermions.
\end{thebibliography}
\end{document}